\documentclass[onecolumn]{aa}

\bibpunct{(}{)}{;}{a}{}{,} 

\usepackage[varg]{txfonts}
\usepackage{gensymb}

\begin{document}

\title{Evidence for a shallow thin magnetic structure and solar dynamo: the driver of torsional oscillations}

\titlerunning{Evidence for a shallow thin magnetic structure and solar dynamo}

\author{T.~R.~Jarboe \and T.~E.~Benedett \and C.~J.~Everson \and C.~J.~Hansen \and A.~C.~Hossack \and \\
K.~D.~Morgan \and B.~A.~Nelson \and J.M. Penna \and D.~A.~Sutherland}

\institute{University of Washington, Seattle Washington, 98195}

\abstract{The solar dynamo and the solar Global internal Magnetic Structure (GMS) appear to be a thin ($\sim$2 Mm thick) structure near ($\sim$1~Mm below) the solar surface. Evidence for these properties are found from the amplitude of the torsional oscillations and in their velocity contours relationship to solar magnetogram; the power to the chromosphere; power to the corona and the solar wind; the current in the helio-current-sheet measured at the radius of the orbit of Earth; the calculated size ($\sim$1~Mm) of the expanding polar flux when it enters the photosphere; the dynamo forces the rigid rotation of the heliosphere with the Sun surface out to at least 1.4 au, giving the solar wind; and from the observation that solar magnetic activity is generated near the surface. A thin stable minimum energy state seems to be covering most of the solar surface just below the photosphere. The magnetic field lines should be parallel to the solar surface and rotate with distance from the surface for 2$\pi$ radians in $\sim$2~Mm. Resistive diffusion helps to push the magnetic fields to the surface and the GMS seems to lose $\pi$ radians every 11 years, causing the observed 180\degree\ flipping of the solar magnetic fields including the flipping of the polar flux. Further evidence for this GMS and its loss is that solar prominences are made of thin sheets of magnetized plasma, which are, likely, remnants of the lost thin sheet of the GMS. The loss process is consistent with the butterfly pattern of the sunspots and with the differences observed between solar maximum and solar minimum in the corona. The solar dynamo drives current parallel to the polar flux, which, in turn, sustains the GMS using cross-field current drive. For completeness, the formation of sunspots, CMEs and flares is discussed.}


\journalname{Draft}
\AANum{Jarboe}

\maketitle

\section{Introduction}

\subsection{History}

A century from the first observations \citep{hale_1908} of the solar magnetic field, consensus still has not been reached \citep{charbonneau_2010} as to the specifics of the mechanism by which that field is generated. That should not be terribly surprising, seeing as the problem is one of defining the inner workings of a tremendous and complicated system that, until the past few decades, could not be directly sampled past the photosphere. 

Research into solar magnetism goes back to George Ellery Hale’s original discovery and quantification, using the Zeeman effect, of the magnetic fields in sunspots. That the Sun would display electromagnetic phenomena was not unexpected at the time, especially as it had already been established that gases would break down into charged particles at the temperature of the Sun \citep{thomson_1903}. Hale’s later work more thoroughly characterized the observed magnetic field of the Sun, including the discovery of such well-known phenomena as the twenty-two year cycle of the magnetic field, and Joy’s Law, that the more easterly part of a magnetically bipolar region is also closer to the equator than the western, trailing part \citep{hale_1919}.

While Hale’s discoveries showed that the magnetic field in sunspots were almost certainly due to the existence of an organized magnetic field structure within the Sun, his observations were of the surface only and could not themselves point the way to whatever precise mechanism produced such a field. Joseph Larmor \citep{larmor_1919} identified flows of conducting fluid, i.e. the solar plasma, as a possible source for the solar magnetic field: a solar dynamo. Larmor’s axisymmetric and equatorially antisymmetric model meshed well with Hale’s observations.

However, in 1933, Thomas Cowling proved \citep{cowling_1933}, in what is now known as his antidynamo theorem, that no steady-state toroidal field could be produced by an axisymmetric dynamo. Thus Larmor’s model was mathematically invalid, and consequently the solar dynamo would have to be more complicated than Larmor’s, if the solar magnetic field was produced by a dynamo at all. Some doubted this, such as Hannes Alfv\'{e}n, who in 1942 proposed that the solar magnetic field was simply a remnant from the formation of the solar system \citep{alfven_1942}.

However, a new instrument - Horace Babcock’s solar magnetograph \citep{babcock_1953} produced new data, particularly that the entire solar field, not just that in sunspots, reversed every eleven years. This rendered Alfv\'{e}n’s model obsolete, and restored the dynamo theory to the lead \citep{stenflo_2015}, for the obstacle discovered by Cowling had been overcome in the meantime, with Walter Elassier’s development of a consistent dynamo theory, which he applied to the magnetic field of Earth \citep{elsasser_1946,elsasser_1955}.

The application of new developments and observations created new dynamo models for the Sun, among them the model proposed by Eugene Parker in 1955, that turbulence plus the Coriolis force could convert toroidal field to poloidal field, while the differential rotation of the Sun plus magnetic buoyancy could convert poloidal field to toroidal field \citep{parker_1955}. This model would come to be systematized into mean-field electrodynamics
\citep{steenbeck_1966}, 
which parameterizes these conversions by decomposing variables into average and fluctuating quantities and solving for the former without directly solving for the latter \citep{radler_2014}, to avoid the problem of solving across the tremendous variation in length scales possessed by a star.

The Parker model was not the only dynamo theory that emerged in this era; Babcock \citep{babcock_1961} and Leighton \citep{leighton_1969} developed a solar dynamo theory of their own, which placed the poloidal field generation near the solar surface. The toroidal field generation, however, still was placed deeper in the Sun, which posed the challenge of how the dynamo could manage to link the two field components’ generation together across the solar convective zone, and this challenge made this model less popular than its alternative \citep{tobias_2002}.

Developments since the 1980s posed challenges to all dynamo models then in existence, from the reevaluation of magnetic buoyancy and diffusivity effects, to the simulation of existing models of the Sun, to the development of helioseismology, which allowed the interior of the Sun to be measured more directly \citep{charbonneau_2010}. While these developments have afforded new opportunities (such as the discovery of the tachocline as a potential location for the solar dynamo), the field of solar physics has not yet settled into a standard model of the solar dynamo \citep{choudhuri_2007}.

Helioseismic measurements have produced \citep{christensen-dalsgaard_2002} an image of the solar interior with a resolution on the order of 1 Mm \citep{hindman_2004}. Features discovered via this technique, such as the tachocline \citep{tobias_2002} and meridional flow have formed key components of modern solar dynamo theories \citep{choudhuri_1995,haber_2002}.

If the magnetic field of the entire Sun (or a thick structure within the Sun) is to change in 11 years, it must diffuse through some mechanism that is far faster than resistive diffusion. Some proposed mechanisms for solving that problem include ambipolar diffusion or turbulent transport \citep{parker_1963}. However, Parker’s later work on the problem says that such mechanisms are not truly applicable to the problem of vector diffusion, and these ``[f]undamendal difficulties with the concept of turbulent diffusion of magnetic fields suggest the solar dynamo needs to be reformulated'' as there is ``no way to account for the [standard] value $\eta \simeq 10^{12}$~cm$^2$/sec, suggesting that it is necessary to rethink the $\alpha \omega$-dynamo for the Sun'' \citep{parker_2009}. Now the situation is not as bleak as Parker describes, with a great deal of research describing the processes which generate this turbulence-driven diffusion \citep{charbonneau_2014} and studies of the role of magnetic reconnection activity in the Sun \citep{yamada_2010}. 

There are considerable data from surface measurements of the solar magnetic fields and solar cycle, but measurements of the internal magnetic fields are limited and the solar dynamo is not fully understood. (Please refer to \citet{hathaway_2010} for an introduction to the solar magnetic cycle and description of measurements that have been made of the solar magnetic field.) While consensus exists that differential rotation is the driver of the solar global magnetic structure (GMS), the location and structure are still active areas of research. Most solar dynamo models \citep{charbonneau_2010,dikpati_1999} assume that the magnetic field is generated at the bottom of solar convection zone, near the tachocline. Alternative models are consistent with the data, such as a dynamo driven by near-surface shear \citep{brandenburg_2005}. However, the model still has the GMS at 35 Mm below the surface, which is much deeper than the $\sim$2 Mm proposed in this paper. It will be shown, to provide the power to the chromosphere and the torque for the torsional oscillations the polar flux and the GMS needs to be about 2 Mm thick. 

\subsection{Requirements of a Global Magnetic Structure model}

Any model of the solar dynamo must therefore be consistent with both the observations made of the Sun and its magnetic field, as well as with the observations made of the behavior of plasmas in terrestrial laboratories. To wit, those criteria include:

First, the solar magnetic structure is global, encompassing the entire body --- as with Earth, the Sun’s overall magnetic field has two opposing magnetic poles, roughly around the poles of rotation --- but only most of the time.

Second, the solar magnetic field reverses itself every eleven years, returning to its original state, more-or-less, after a period of twenty-two years in total \citep{babcock_1961}.

Third, the solar magnetic field is correlated with sunspot activity, i.e., surface phenomena, which suggests that there should be a close link between the solar surface and the region of magnetic field generation. 

Fourth, solar prominences are made of thin sheets of magnetized plasma and the solar corona changes its shape during the solar cycle.

Fifth, torsional oscillations are strongest near the surface \citep{spruit_2003} and follow the solar cycle. They also exist during solar minimum. 

Sixth, the fast solar wind accelerates as it expands.

\subsection{Proposed Solution}

Consideration of the aforementioned constraints from solar observations and experimental exploration into plasma physics suggest that a thin layer near the surface ($R\simeq 0.9986 R_o$) should be considered as the source of the external solar magnetic field. 

First, the polar flux reverses entirely every eleven years; in other plasma physics problems, a full field reversal in a conducting medium is considered highly unlikely to happen on length scales much greater than the characteristic skin depth for that material and that timescale. This would suggest that the GMS ought not to be much farther into the Sun than an eleven-year skin depth below the photosphere. 

Second, studies of the surface and subsurface of areas from where magnetic structures emerge have shown there is no indication of any movement to the area before the emergence \citep{birch_2013}, which would suggest either that the structures emerge at truly breakneck speeds such that the motion outpaces the time resolution of the studies, or, more plausibly, that there is very little motion at all, which would indicate that the magnetic field has its source near the surface. 

Third, the polar flux has a thin cusp source giving the localized corona \citep{amenomori_2013}.

Fourth, the recent approximate halving of the dynamo power, in cycle 24, lowers the solar irradiance by the order of 0.02\% (an estimate based on Fig.~\ref{figure_1} of \citep{kopp_2016}).  Assuming that the change in irradiance and dynamo power are directly connected, that would mean the dynamo power was of order 0.04\% of the irradiance ($1.2 \times 10^{23}$~W of dynamo power with large uncertainly) in 1998 when the signed polar flux was $3 \times 10^{14}$~Wb \citep{jiang_2011}, it will be shown that the gradient scale length of the magnetic structure was of order 0.17 Mm, a thin structure. 

Fifth, rapidly changing magnetic activity in the photosphere is attributed to a ``magnetic carpet'' near the surface \citep{savage_1997}.

Sixth, torsional oscillations are strongest near the surface and are correlated with magnetic activity.

Therefore, unlike most models \citep{charbonneau_2010} this paper explores the possibility that the internal solar magnetic field is near the surface, specifically in the supergranulation region related to the magnetic carpet \citep{priest_2002} below the photosphere. The structure is thin and close to the surface, such that the magnetic decay time is comparable to the solar cycle period. Differential solar rotation in the internal polar flux drives the solar dynamo, which sustains a stable equilibrium and powers the chromosphere and corona. The model can explain butterfly diagrams 
\citep{maunder_1904},
flipping of the polarity of sunspots \citep{hale_1919}, and the flipping of the polar magnetic field \citep{babcock_1961}. The model also explains a relationship between velocity contours of the torsional oscillations \citep{howe_2009} and solar magnetograms \citep{hathaway_2010}.

The main argument against near-surface dynamos has been that turbulence is too strong to allow for the formation of large-scale magnetic structures. It will be shown herein that with a magnetic structure near the minimum energy state, turbulence does not destroy the structure nor helicity. A GMS cannot exist in the convective zone because of severe turbulence and cannot exist in the photosphere because of high resistivity and low density. However it appears to exist in the thin region between these regions where the conditions are right for magnetic self-organization to have a significant effect on plasma behavior. Assuming the 9-day period observed in some solar activity \citep{verkhoglyadova_2011,temmer_2007} is the toroidal Alfv\'{e}n time of the GMS, then $B_o$ is about 0.5~T. With turbulent velocity v = 20 m/s; the density $\rho$ = 0.002 kg/m$^3$; the resistivity $\eta = 6 \times 10^{-5}$~$\Omega$m and the current density j = 1.6 A/m$^{2}$; the magnetic energy density is over 10$^5$ greater than $\rho v^2$ and $\bf v \times B$ is over 10$^5$ greater than $\eta j$. Thus the magnetic field will not be destroyed by the turbulence and the dynamo term can dominate the resistive term in sustaining and stabilizing the equilibrium.

\section{Derivations and observations of the self-organization of solar magnetic structures}

Self-organization of the solar plasma allow the formation and sustainment of a shallow, thin Global Magnetic Structure (GMS). The Woltjer-Taylor minimum energy principle, helicity injection current drive, and the scaling of helicity conserving dynamics will be discussed. The application of these concepts to the solar dynamo is the new physics given by this paper. Demonstrating that the complex solar magnetized plasma dynamics can be understood by the proper application of a few concepts is a major advance in plasma physics and solar physics.

\subsection{The minimum energy principle}

The Woltjer-Taylor minimum energy principle states that magnetic plasma relaxes toward a state of minimum energy while conserving helicity (Taylor, 1986). The MECH state is stable. Velocity turbulence does not dissipate helicity. In the solar plasma just below the photosphere, the collision frequency is higher than the cyclotron frequency, and the plasma has the properties of liquid metal. Therefore, it is well described by resistive magnetohydrodynamics with a generalized Ohm’s law, which is found by a Lorentz transformation from the plasma frame where ${\bf E} = \eta{\bf j}$. That generalized Ohm’s law is:	

\begin{equation}
{\bf E} = - {\bf v \times B} + \eta {\bf j}
\label{equation_1}
\end{equation}

\noindent
where ${\bf v}$ is the plasma velocity, $\eta$ is the resistivity and ${\bf j}$ is the current density. Magnetic helicity is the linkage of magnetic flux with magnetic flux \citep{moffatt_1978}, $K = \int {\bf A \cdot B}\, \mathrm{dvol}$, where ${\bf A}$ is the vector potential and B is the magnetic field, and helicity is dissipated by ${\bf 2 E \cdot B}$ (Finn and Antonsen, 1985). While velocity turbulence cannot dissipate helicity (as the velocity-dependent term in E is in a term perpendicular to ${\bf B}$ due to the cross product,) it can dissipate excess magnetic energy by ${\bf j\cdot E}$, which is a more rapid process than helicity dissipation. This would drive the magnetic structure to the MECH state. In turbulent regions of the Sun, due to turbulence’s continual action, which dissipates only the excess magnetic energy but does not affect magnetic helicity, the magnetic structure can exist in the MECH state, and perhaps only in or near the MECH state. Furthermore, helicity decays on the resistive time scale (the longest characteristic time scale in the system \citep{edenstrasser_1995}) and the MECH state is stable \citep{edenstrasser_1995}. Therefore, MECH magnetic structures may be tolerant to turbulence. 

Complete relaxation, with magnetic flux boundary conditions, gives ${\bf \nabla \times B} = \lambda_{eq} {\bf B}$, where $\lambda_{eq}$ is a global constant \citep{woltjer_1958,taylor_1986}. Utilizing Ampere’s law yields $\lambda_{eq}= \mu_0 j/B$, where $\lambda_{eq}$ is inversely proportional to the characteristic size of the system. The larger class of force-free states obey ${\bf \nabla \times B} = \lambda {\bf B}$ where $\lambda$ is not a global constant but is still the rate of rotation of the magnetic field for motion across the magnetic field.

\subsection{Sustaining stable plasma currents by helicity injection current drive}

Sustained plasmas can have an arbitrary $\lambda$ value as part of the boundary conditions (defined by an external circuit), and the MECH state does not have uniform $\lambda$ \citep{jarboe_1994}. Ideally the injector fields will have much of its boundary parallel to the rest of the equilibrium and separatrix can form between the region with $\lambda_{\rm inj}$ defined by the boundary conditions and the region where $\lambda_{\rm GMS}$ is found by this self-organization. The last ingredient is perturbations that cause an anomalous viscosity in the electron fluid giving the cross-field current drive that can sustain a stable plasma. \citep{jarboe_2015,hossack_2017} The perturbations can be imposed by the external geometry or circuit or produced by plasma instability. Since the energy per unit helicity is $\lambda / 2 \mu_0$ helicity flows from high-$\lambda$ to lower-$\lambda$ regions.  The helicity injection circuit maintains the higher $\lambda_{\rm inj}$. This leads to a two-step-like $\lambda$-profile of the sort observed in the HIT-SI experiment \citep{jarboe_2012,victor_2014}. 

The Lundquist number, $S$, defined as the resistive diffusion time normalized by the toroidal Alfv\'{e}n time is the same order in helicity injection current drive experiments as in the Sun. This Lundquist number characterizes resistive MHD in toroidal geometry. The measurements and simulations of HIT-SI plus the simplicity of solar current drive and the excellent solar data give a clear picture of the sustainment of stable equilibrium.

\subsection{Scaling of helicity conserving dynamics:}

As shown in Sect.~\ref{sec_2.4}, the magnetic diffusion time of the GMS is of order 10 years, which is much longer than the expansion time of the plasma from the solar surface to the Earth.
Because the diffusion time increases as the size squared, it increases faster than the time of the expansion.
Thus, the flux and helicity are both always conserved. 
Note: 
\begin{eqnarray}
\frac{B^2 s^3}{B^2 s^4} \propto \frac{\text{Energy}}{\text{Helicity}} \propto \frac{1}{s} ; \frac{Bs}{Bs^2} \propto \frac{\text{Current}}{\text{Flux}} \propto \frac{1}{s} \label{equation_2}
\end{eqnarray}
where $s$ is the characteristic size ($\equiv \pi / \lambda \simeq$ distance from the solar source of the plasma).

For a self-similarly, steady-state expanding magnetic structure conserving helicity and flux, the helicity passing through the surface a distance $s$ from the source is the same for all $s$.
Thus, from Eq.~\ref{equation_2}:
\begin{eqnarray}
\text{Power} \propto \frac{1}{s} ; P_s s = \text{const} \label{equation_3}
\end{eqnarray}
where $P_s$ is the power through the surface at $s$.
Also, the magnetic flux penetrating the surface of distance $s$ from the source is the same for all $s$.
Thus, from Eq.~\ref{equation_2}
\begin{eqnarray}
\text{Current} \propto \frac{1}{s} ; I_s s = I s_{\rm inj} \label{equation_4}
\end{eqnarray}
where $I_s$ is the current through the surface at $s$.
Define $P_{\rm dym}$ as the power through the surface when the expansion begins (at the exit from the GMS) in one hemisphere.
Thus,
\begin{eqnarray}
P_{dym} s_{\rm inj} = \frac{V_{\Delta} I \pi}{\lambda_{\rm inj}} = \frac{V_{\Delta} I \pi}{\frac{\mu_0 I}{\psi}} = \frac{V_{\Delta} \psi \pi}{\mu_0} \label{equation_5}
\end{eqnarray}
where $V_{\Delta}$ is the dynamo voltage drop in one hemisphere.
The power into a region is:
\begin{eqnarray}
P_{\text{region}} = \frac{V_{\Delta} \psi \pi}{\mu_0} \left( \frac{1}{s_{en}} - \frac{1}{s_{ex}} \right) \label{equation_6}
\end{eqnarray}
where $s_{en}$ and $s_{ex}$ are the sizes of entering and exiting the region.

\subsection{Resistive diffusion leading to surface magnetic activity:}
\label{sec_2.4}

\begin{figure}[!htbp]
\includegraphics[width=\linewidth]{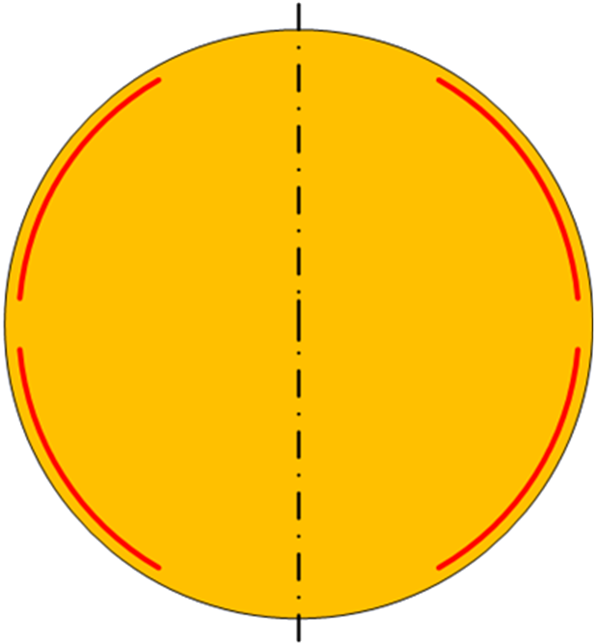}
\caption{\label{figure_1}Cross section of the Sun and of the thin, stable, toroidally symmetric global magnetic structure (GMS) under the solar surface in red.}
\end{figure}

\begin{figure}[!htbp]
\centering
\includegraphics[width=0.7\linewidth ]{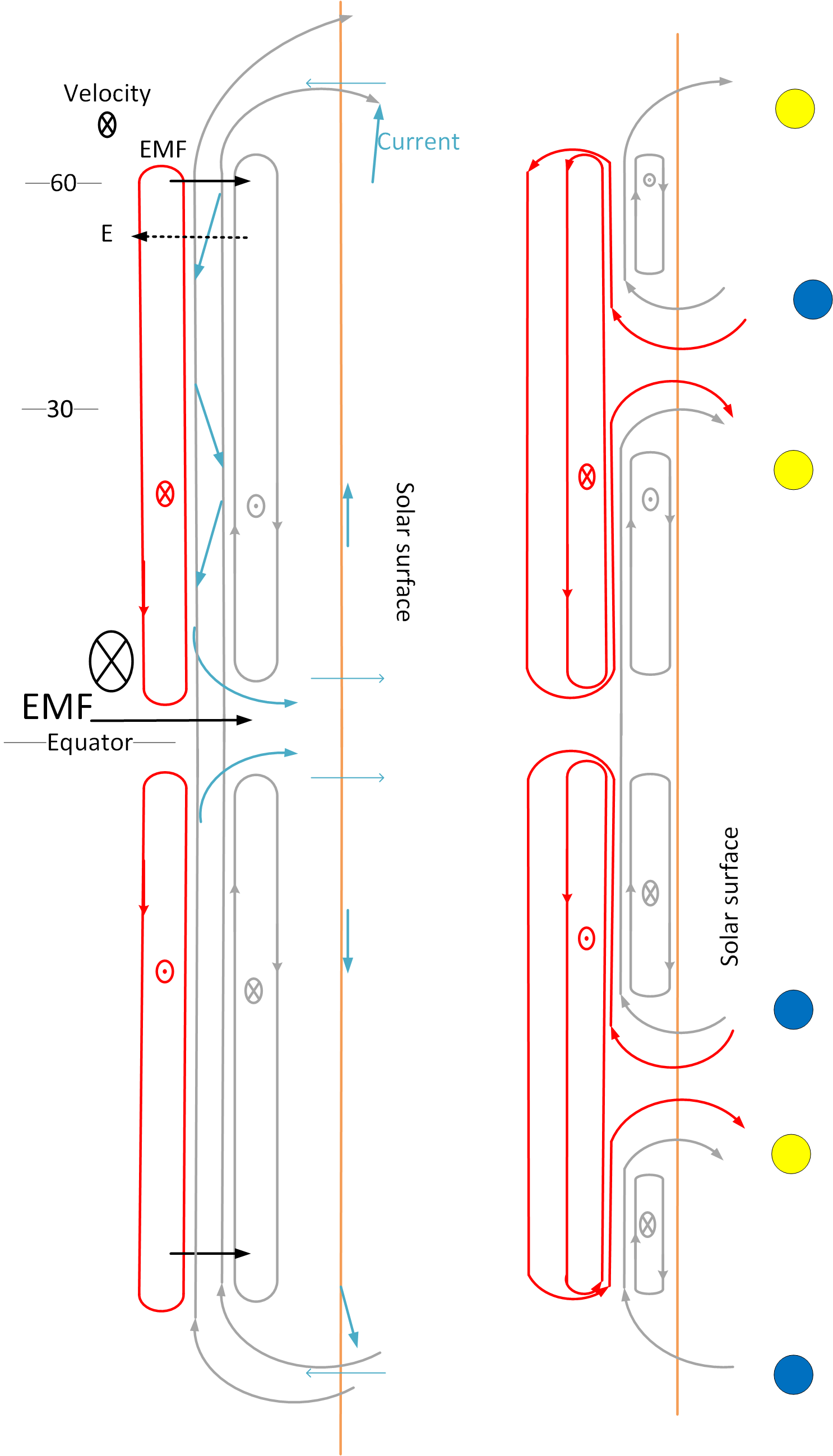}
\caption{\label{figure_2} The figures represent snapshots in time of poloidal cross sections of the GMS. Based on the timeline defined by the figures, the approximate time of events are given at the bottom. a) Current directions are based on the acceleration in Fig.~\ref{figure_5}b) and are shown as blue arrows. Colored and gray closed structures are called lobes. Arrows on the lobe are magnetic field direction. Black arrows are electric field (dashed) and EMF (solid). The black direction X is for the surface velocity. b) The colored circles give the direction of the magnetic field of Fig.\ref{figure_5}a) at that position, yellow is out of the surface blue is into the surface. Figure on left is the year 1997 on the right is about 1999.}
\end{figure}

Assuming the GMS is thin and close to the surface, its helicity loss time is $\sim$11 years. Helicity injection \citep{taylor_1986} from the solar dynamo, as discussed below, sustains it, and instability might provide the perturbations needed for sustainment and for keeping it in the MECH state as discussed above.The GMS appears to consist of two thin (each $\sim$2 Mm thick), MECH states, near the surface of the Sun. As shown in Fig.~\ref{figure_1} each covers the Sun from 15 degrees, or less, to about 60 degrees latitude with large-scale longitudinal symmetry. One MECH state is in the northern hemisphere and is coupled by common polar magnetic flux to one in the southern hemisphere. If the GMS exists as a MECH state, then $\boldsymbol{\nabla} \times \mathbf{B} = \lambda_{\rm GMS} \mathbf{B}$ is dominated by the rapid spatial variation in the radial direction, and variations in the other directions can be ignored. Thus, if the GMS is in a thin layer it will be approximately a MECH state with the only relevant variation being in one dimension. Solving $\boldsymbol{\nabla} \times \mathbf{B} = \lambda \mathbf{B}$ gives the 1D MECH state with the longitudinal and latitudinal magnetic field as $B_{\rm long} = B_o\sin \left( \lambda_{\rm GMS}\Delta R \right)$, $B_{\rm lat} = B_o\cos \left( \lambda_{\rm GMS}\Delta R \right)$, where $\Delta R$  is the radial position and is zero at the inside edge of the GMS. The polar flux, with parallel current driven by the solar dynamo, threads between two lobes of approximately $\pi$ rotation each of the magnetic field as shown in Fig.~\ref{figure_2}a). The figure has a greatly expanded radial direction and compressed meridional direction, and the solar surface is shown as straight for convenience. Arrows along streamlines and red and gray into and out-of the page directions are for the $\mathbf{B}$-field. The closed structures are toroidally symmetric lobes of spatially-rotating magnetic field with purely toroidal field at the center and mostly poloidal field at the edges. The northern hemisphere has negative helicity ($\mathbf{j}$ anti-parallel to $\mathbf{B}$) and negative $\lambda_{\rm GMS}$, and the southern hemisphere has positive helicity and positive $\lambda_{\rm GMS}$. Every 11 years $\pi$ radians of rotation for each state escape, causing the observed magnetic activity with the magnetic fields flipping 180\degree. The driven polar flux sustains the GMS as discussed in Section 2.2.

Our present understanding of self-organization leaves little doubt that a thin dynamo will have the MECH state structure and that 180\degree\ flipping of the solar magnetic field and the surface magnetic fields comes from losing $\pi$ radians of the structure each 11 years. That torsional oscillations (discussed below) are strongest near the surface and exist during solar minimum, strongly confirm the thin shallow dynamo.

Buoyancy and Rayleigh-Taylor stability considerations do not preclude the existence of the GMS. The condition for neutral buoyancy is that the density inside and outside of the GMS are equal and the total pressures $(p + \frac{B^2}{2\mu_o})$ are equal at a given radius. A cooler plasma inside the GMS allows both conditions to be met. The higher ratio of specific heat ($\gamma$) of magnetic pressure and the increase of heating deeper in the Sun may stabilize buoyancy forces. The equilibrium demands \citep{spitzer_1962} $0 = -\mathbf{\nabla}p + \rho\mathbf{g} + \mathbf{}\frac{B^2}{2\mu_o}$ where $\rho$ and $\mathbf{g}$ are the density and the gravity and $\mathbf{j} \times \mathbf{B} = \frac{\mathbf{\nabla}B^2}{2\mu_o}$. Rayleigh-Taylor stability requires $\mathbf{\nabla}\rho$ be negative. In this region $\frac{\mathbf{\nabla}T}{T} \ll \frac{\mathbf{\nabla}\rho}{\rho}$ and $\rho\mathbf{g} \gg \frac{\mathbf{\nabla}B^2}{2\mu_o}$ gives the requirement for negative $\mathbf{\nabla}\rho$. For a gradient scale length of about 0.1 Mm at 2 Mm below the surface requires $B_o$ be less than about 0.7 T. In addition, a stable GMS might stabilize the Rayleigh-Taylor instability. The poloidal flux of the two lobes is about 1.5 times the polar flux, giving a flux ``amplification'' easily achieved on HIT-SI \citep{victor_2014}. Thus, the magnitude of $\mathbf{B}$ can have gradient scale length of order $0.1s_{\rm GMS}$ at the upper and lower boundaries allowing most of the volume to have a uniform $B_o$.

This 1D MECH state has a uniform magnitude magnetic field, $B_o$. Since the energy per unit helicity is proportional to $\lambda_{\rm GMS}$, its smallest value that satisfies the constraints defines the minimum energy state and $\lambda_{\rm GMS} \leq \lambda_{\rm inj}$. The lobe height, $s_{\rm GMS}$ defined as $\pi$ rotation of $\mathbf{B}$, is $\frac{\pi}{\lambda_{\rm GMS}}$. The value of $B_o$ adjusts to match the helicity content (injected minus dissipated) of the GMS. While $B_o$ may be fairly constant over much of the GMS volume,
$\mathbf{B}$ must go to zero at the boundary.

The details of the GMS structure and the details of the loss of the upper lobe on the 11 year cycle are not known and more data and computer modeling are needed for a quantitative picture. The effects of gravity and buoyancy and transmission of the solar power will need to be included. However, how resistivity gradients might help movement to the surface is now discussed without considering other motion drivers. The following assumption are made: 1) The lobe boundaries move by resistive diffusion. 2) The GMS is a 1D MECH state stabilized by low-S and local instability and/or high turbulence keeping it near the stable MECH state. 3) Dynamo flux, with $\lambda_{\rm inj}$, threads two resistively-expanding lobed structures and gives them equal helicity injection. Two lobes are required so that helicity can be injected without changing the global toroidal flux or global current. The dynamo flux must pass between the lobes so that the dynamo flux links the flux of the upper lobe, giving the maximum helicity. 4) Relaxation fills the resistively-produced new volume with GMS, keeping the boundary thin.  The boundary of thickness $\delta \simeq 0.1s_{\rm GMS}$ separates changes in $\mathbf{B}$, which preserves the $(p + \frac{B^2}{2\mu_o})$ jump condition. 5) Each lobe is a $\pi$-radians high 1D MECH state.

 The two lobes are separated by the dynamo flux. The other boundaries with the unmagnetized plasma then move with velocity:

\begin{equation}
v = \frac{\eta}{\mu_o\delta}
\end{equation}

Resistivity decreases with depth into the Sun so the top surface moves toward the solar surface faster than the bottom surface moves away, giving a total motion up, towards the solar surface.

\begin{figure}[!htbp]
\includegraphics[width=\linewidth ]{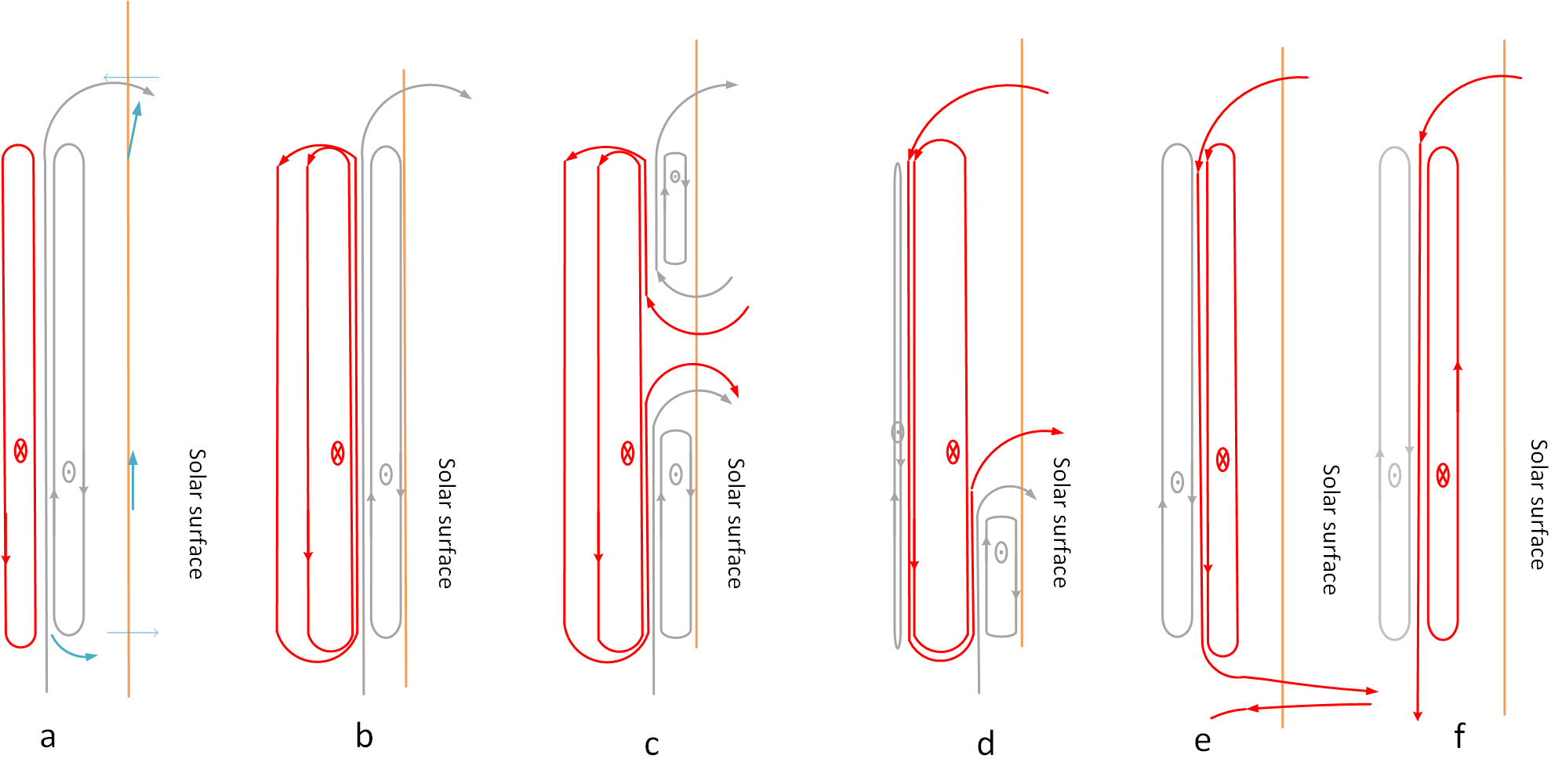}
\caption{\label{figure_3} More detailed-in-time depiction of the process of the polar flux reversal (shown in Fig.~\ref{figure_2}) for a single hemisphere only, displaying the formation of new lobes, the displacement of old lobes, and the resistive process by which the polarity reverses. a) The MECH state near solar minimum. (The pattern of the radial field at the solar surface is seen at 1997 in Fig.~\ref{figure_5}a) Resistive diffusion moves the boundaries. (1997.3) c) The GMS becomes too close to the surface for a stable equilibrium and breaks through the surface, breaking the old lobes in two. (1999) d) Meridional flow moves the polar side to the polar region but in the equator side the flow is countered by tension in the polar flux resulting in a slow movement toward the equator. The new polar flux is being dynamo driven and a new lobe starts growing from the helicity injection (1999.5) e) Old upper lobe has left allowing the new polar flux to connect with its counterpart in the in the southern hemisphere to form the new polar flux. (2005) f) The new MECH state near solar minimum. (2009) The polar flux came from the red flux. }
\end{figure}

Figure~\ref{figure_3} depicts the process of the polar flux reversal. This is for a single hemisphere only, displaying the formation of new lobes, the displacement of old lobes, and the process by which the structure reverses polarity. This half-cycle shows the unwrapping of flux from around what becomes the new upward-directed lobe. The flux becomes the new reversed polar flux.

When the dynamo drive starts up on the new polar flux, which was part of the GMS, the pinch effect may compress this flux and causes the bottom to move up by the amount it moves down in the rest of the cycle. See Fig.~\ref{figure_3}d). Similarly, the loss of the upper lobe causes the top to move down by the amount it moves up in the rest of the cycle. The size of the red lobe may become the width to make these motions equal. The time of the 11-year cycle is the time it takes the upper edge of the red lobe in Fig.~\ref{figure_3}d) to reach the level at which the upper lobe is released. Assuming the stable GMS keeps the equilibrium stable, the lobe leaves from insufficient mass above it to keep it submerged.

Using Spitzer resistivity \citep{spitzer_1962} for the plasma plus a constant-cross-section $(5.7 \times 10-19 m^2)$ correction for electron collisions with neutrals \citep{kolesnikov_1962}, the Saha equation for percentage ionized and estimates of solar properties \citep{christensen-dalsgaard_1996}, the resistivity in $\Omega$m as a function of depth below the photosphere $x$ in meters ($x = 0$ at the surface) is approximately:

\begin{equation}
\eta = \frac{60}{x + 8000}
\end{equation}

With this resistivity, the front of the lobe must start at $x = 0.57 Mm$ to reach the surface in 11 years. The time for the lobe to diffuse to the surface is of order 11 years fairly independently of the value of $B_o$. The conditions that the three interfaces must have the same period determines the size, the depth and the period of the GMS. This might also remain true when a more complete model of the dynamo dynamics in the Sun is given. This region of the Sun is not too multi-scale for computers to model quantitatively and resistive MHD is very accurate plasma physics for this problem. A contribution of this paper is to show where to look for the dynamo.

The polar flux connects the northern and southern halves of the GMS, below the solar surface across the equator, synchronizing their magnetic activities, and possibly balancing meridional forces. Most importantly, this flux crosses the differential rotation of the solar surface, producing the voltage which drives current that sustains the GMS and other solar activity. The high turbulence of the solar convection can produce large enough perturbations to give the cross-field current drive that assures a uniform $\lambda$ over each lobe, and assures a grossly stable sustained GMS. The GMS can stably form and grow in place, a lobe on each side of the driven current sheet, without the need of any global instability, otherwise associated with relaxation, and without changing the total flux or current. Such instability would have made the survival of such a thin GMS doubtful. In addition, the polar flux has the minimum energy when it is purely poloidal and the magnetic forces are greater than the viscous forces. Thus, the internal polar flux is assumed approximately purely poloidal.

\section{Results: the self-organization of solar magnetic structures}

\begin{figure}
\center{\includegraphics[width=\linewidth]{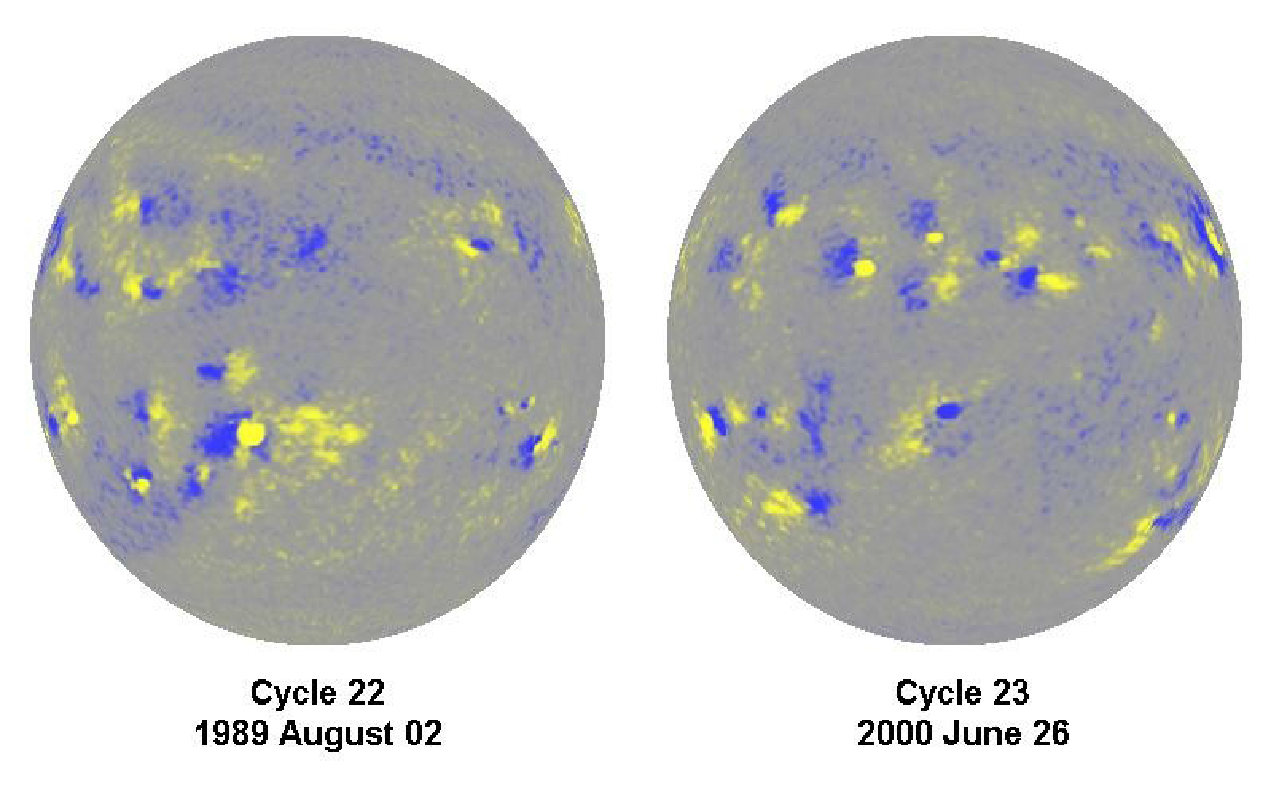}}
\caption{“Hale’s Polarity Laws. A magnetogram from sunspot cycle 22 (1989 August 2) is shown on the left with yellow denoting positive polarity and blue denoting negative polarity. A corresponding magnetogram from sunspot cycle 23 (2000 June 26) is shown on the right. Leading spots in one hemisphere have opposite magnetic polarity to those in the other hemisphere and the polarities flip from one cycle to the next.” \citep{hathaway_2010} The 1989 magnetogram will be similar to 2010 in for comparing to Fig.~\ref{figure_2}.}
\label{figure_4}
\end{figure}

\begin{figure}
\centering
\center{\includegraphics[width=0.65\linewidth]{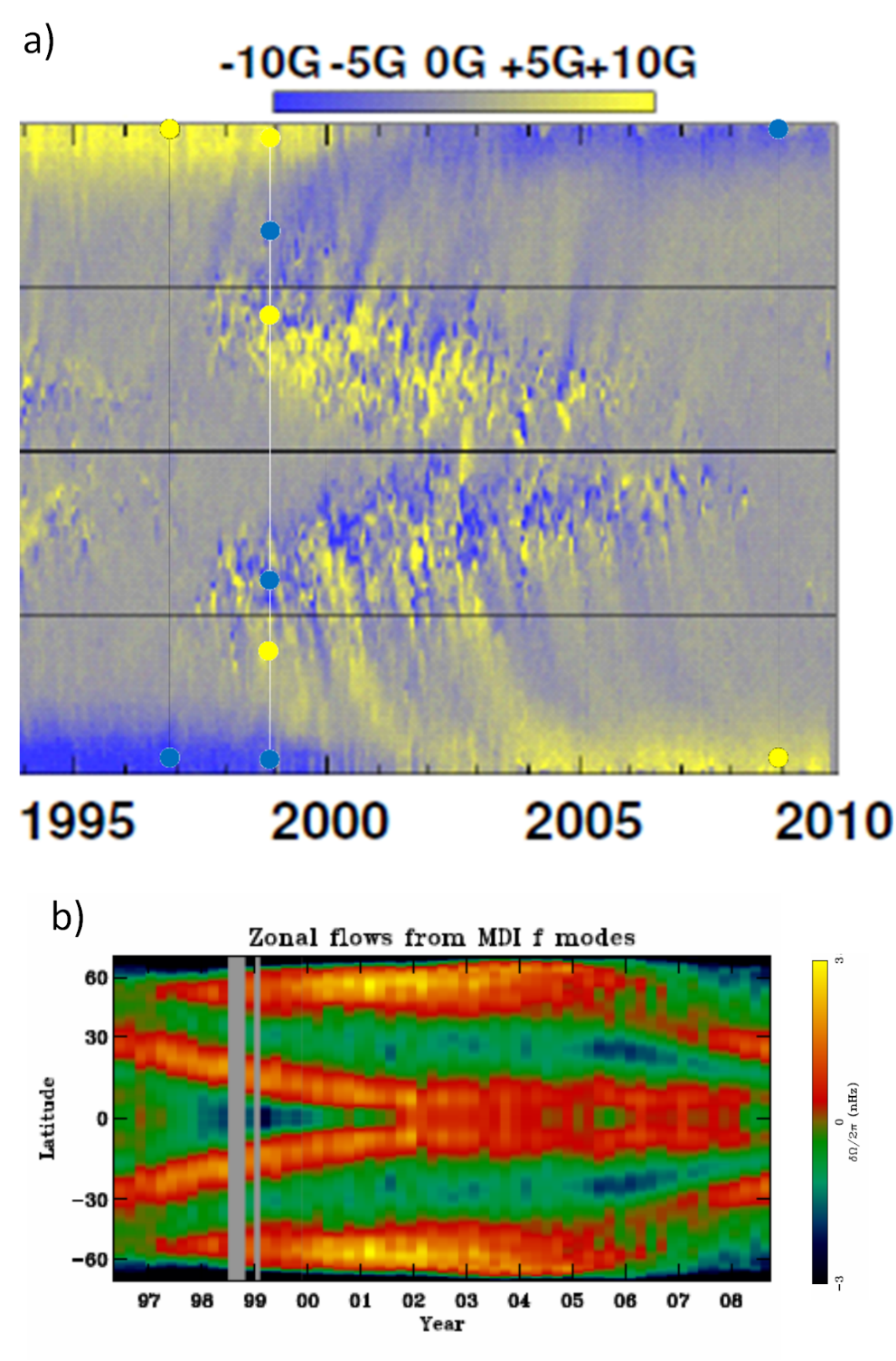}}
\caption{a) “A Magnetic Butterfly Diagram constructed from the longitudinally averaged radial magnetic field obtained from instruments on Kitt Peak and SOHO. This illustrates Hale’s Polarity Laws, Joy’s Law, polar field reversals, and the transport of higher latitude magnetic field elements toward the poles.” \citep{hathaway_2010} The first two vertical lines and the colored circles correspond to the two times sketched in Fig.~\ref{figure_2}.  b) The velocity contours of the torsional oscillations from \citep{howe_2009}. }
\label{figure_5}
\end{figure}

Figures \ref{figure_4} and \ref{figure_5}a) are solar data, with figure captions, from \citep{hathaway_2010}. Assuming that toroidal average of the radial component of the solar magnetic field of Fig.~\ref{figure_5}a) is from poloidal field escaping the surface, and the sunspots’ toroidal fields are in the direction of the magnetic axis of the escaping upper lobe, then the polar flipping and the sunspot pattern agree with coupled 1D MECH states that lose one lobe or $\pi$ rotation in $\bf B$ every 11 years. In going from Fig.~\ref{figure_3}a) to f), the lobe near the surface is lost. The into-the-page direction on Figs.~\ref{figure_2} and \ref{figure_3} is from left to right on Fig.~\ref{figure_4}. The direction of the toroidal magnetic field of the sunspots during the time from that of Fig.~\ref{figure_3}a) to that of Fig.~\ref{figure_3}d) are the same as that of the lost lobe. Thus, the directions will be flipped for the next butterfly pattern as observed.

Figure~\ref{figure_5}a) shows that most of the activity is below 30\degree, yet the polar flux is below the surface to 60\degree\ at the solar minima, consistent with the first surfacing of the upper lobe at 30\degree\ latitude. At year $1997.5$ in Fig.~\ref{figure_5}a), this event seems to trigger the beginning of the solar activity at 30\degree\ latitude. The activity then propagates towards the equator, giving the butterfly pattern. Figure \ref{figure_2}b) shows the internal structure after the old polar flux reaches the surface. The radial magnetic field pattern shown agrees with Fig.~\ref{figure_5}a) at year 1999. The model captures the five changes in direction of the radial field in going from top to bottom. The yellow and blue colored circles positions are the same in Fig.~\ref{figure_2}b) and Fig.~\ref{figure_5}a) at 1999. The radial magnetic fields of Fig.~\ref{figure_3}a) and \ref{figure_3}f) agree with Fig.~\ref{figure_5}a) at the solar minimums before and after the year 2000, respectively. (For this paper, numerical examples refer to this cycle.)

It appears that the resistive diffusion leads to new flux penetrating the surface, as shown in Fig.~\ref{figure_3}c). In Fig.~\ref{figure_3}d) the higher latitudes of the new penetrations move to the poles where some (gray) is releases from the Sun and some (red) becomes the new polar flux, anchored by the new red lobe. The lower latitudes of the new penetrations slowly move (allowing time for sunspots) to the equator releasing the new polar flux from the solar surface and releasing the old polar flux from the Sun due to the loss of its gray anchor. The high latitude streaks to the poles and the coloration at low latitudes in Fig.~\ref{figure_5}a) are consistent with this description. The gray lobe breaking off and floating to the surface provides the thin sheets of magnetized plasma for prominences and the material of other magnetic activity of the cycle. 

\section{Result: Two thin rings may provide the power and flux for the heliosphere and the torque for torsional oscillations and the fast solar wind}

Figure~\ref{figure_2}a) shows the current driving mechanism.  In resistive MHD, the generalized Ohm's law (Eq.~\ref{equation_1}) has an additional term if the plasma is moving at velocity $\bf{v}$.  The $\bf{v}\times\bf{B}$ term has a minus sign because the electric field, like in a battery, results from a charge separation response to an electromotive force (EMF) in the opposite direction.  The charge separation creates an electrostatic potential, $V$ that drives the rest of the plasma.  The dynamo driven current that flows against the dynamo $\bf{E}=-\bf{v}\times\bf{B}$ and flows with the dynamo EMF.  Since the plasma resistance is small, $V$ is approximately a flux surface quantity. The current that flows across the flux surface in the EMF direction equals the current that flows across parallel to $\bf{E}$. $V$ is determined by this balance.

The dynamo EMF dominates at the high velocity locations under the solar surface where the velocity is determined by the powerful solar hydrodynamics. In the heliosphere the velocity is driven by $V$ and the EMF is a back EMF and smaller than $V$, resulting in current in the $\bf{E}$ direction. The EMF dominates at the high-velocity equator region and $V$ dominates at the lower velocity poles. In general, $V$ dominates where the polar flux comes to the surface as in a flare and the EMF dominates at the highest velocity regions under the surface. The direction of the crossing current determines the direction of the acceleration force on the surface.

\begin{figure}
\includegraphics[width=\linewidth]{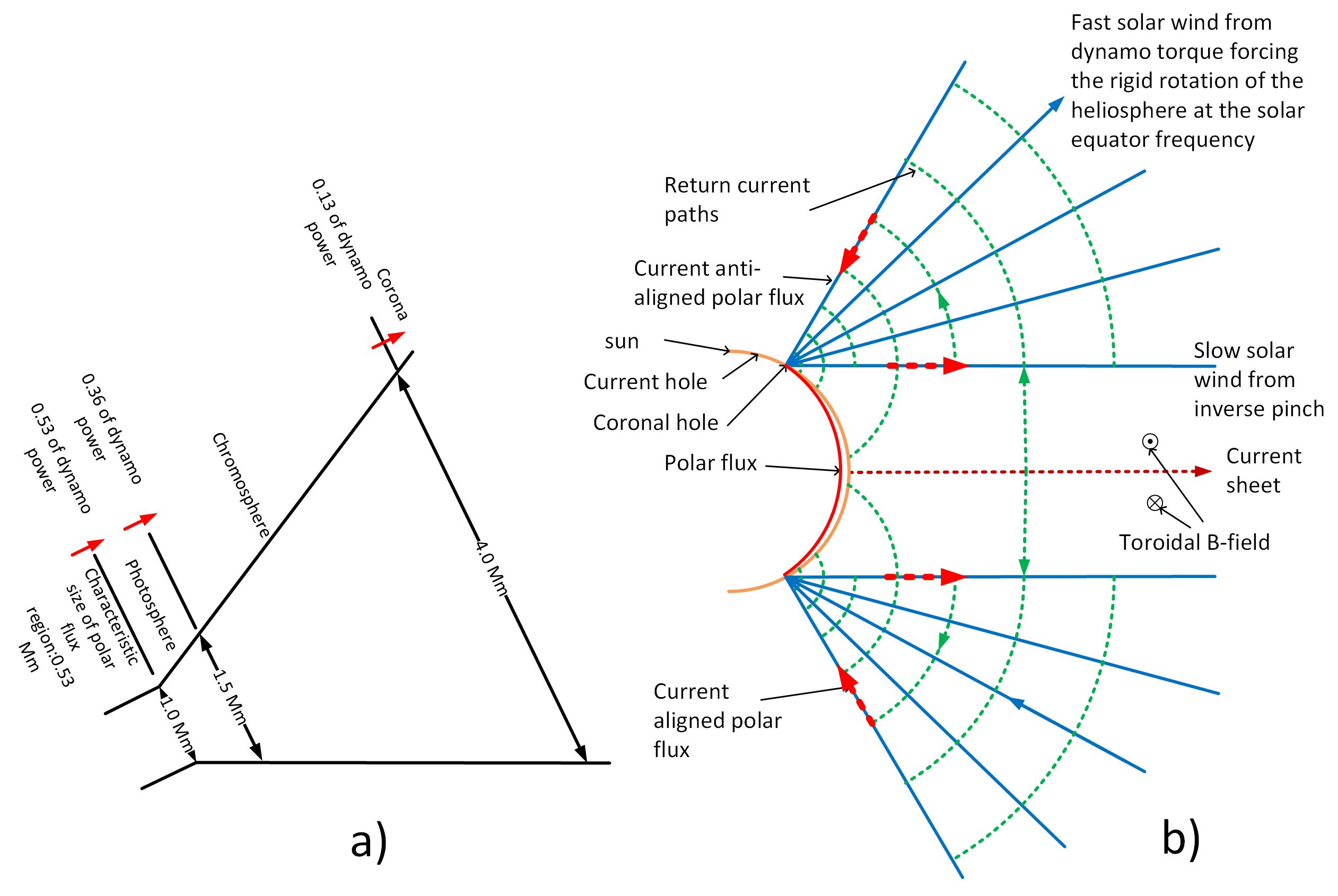}
\caption{Simplified picture of the expansion of the driven polar flux into the solar atmosphere at solar minimum showing the corona, coronal holes, current holes, current sheet, and solar wind. a) Small scale expansion showing power flow. b) Larger scale expansion showing the component of current and magnetic field stream lines in the poloidal plane. Solid lines represent the polar flux, which has current parallel to it in the Southern Hemisphere and anti-parallel to it in the Northern Hemisphere. Dotted paths are of the components of current of the solar circuit with green for cross-field current and red is field-aligned. The heavy red arrows show the current that drives the fast solar wind. }
\label{figure_6}
\end{figure}

The EMF dominated current decelerates the surface and the $V$ dominated current accelerates the surface, lowering the differential flow, which powers the solar dynamo. (If the rotation frequency were identical everywhere the EMF would be uniform and be exactly canceled by $V$, resulting in no current being driven.) This variation in the direction of crossings determines the velocity pattern of the torsional oscillations. See Fig.~\ref{figure_5}b). Thus, at the time of Fig.~\ref{figure_2}a) the driven current flows into the GMS at the poles and out at the equator.  See the blue arrows in Fig.~\ref{figure_2}a). Note that the current is aligned with the field in the Southern Hemisphere and anti-aligned in the Northern Hemisphere as is required to match the data. When the open poloidal flux and current leave the GMS, they rapidly expand while conserving helicity and losing magnetic energy to the plasma through heating and acceleration, which could power the chromosphere, the corona, where there is no polar flux, the solar wind, as in Fig.~\ref{figure_6}.  This expansion into the heliosphere dominates the dynamo power loss.

Using Eq. \ref{equation_1} and Fig.~\ref{figure_2}a), and the center of the Sun as the rest frame, the EMF voltage is $\int vB \mathrm{d}R$, the flux, $\psi$, is $2\pi R_{\rm axis}\int B\mathrm{d}R$: so the EMF is $\psi \omega / 2\pi$, where $R_{\rm axis}$ is the distance from the surface to the axis of rotation, $v$ is the surface velocity, and $\omega$ is $v/R_{\rm axis}$.  Thus, the voltage source is the same as in many earlier solar dynamo models of the magnetic fields.  The voltage drop in the solar circuit is the difference between the highest and lowest EMF and $\psi \omega / f 2 \pi$, is used for the voltage drop with $f$ to be found.
Assuming $\lambda_{\rm inj}$ is uniform the torque about the solar axis is $\Gamma = R_{\rm axis} \int {\bf I}_{\rm cross} \times {\bf B} \mathrm{d} R$ where ${\bf I}_{\rm cross}$ is the current crossing the flux surface at $R$ and equals $\lambda_{\rm inj} \psi / \mu_0$ at $R=R_o$ and equals zero on the inside edge of the polar flux so ${\bf I}_{\rm cross}=\lambda_{\rm inj} \psi^\prime /\mu_0$ where $\psi^\prime \left( R \right)$ is the flux as a function of $R$.  Now $\Gamma=\int \lambda_{\rm inj} \psi^\prime \mathrm{d} \psi^\prime /2\pi \mu_0$ or $\Gamma=\psi I / 4\pi$.  The heliosphere dominates the impedance which sets $\lambda_{\rm inj}$ ($= \omega \mu_0/2\pi Z$ where $Z = V_\Delta / I$).

The highest angular velocity on the solar surface is at the equator, with the angular velocity almost uniform from $0^\circ$ to $15^\circ$ latitude \citep{christensen-dalsgaard_2007}.  At the equator, the rotational frequency is given by $\omega/2\pi = 460$ nHz.  It is believed that $2.65 \times 10^{22}$ W are needed to heat the whole chromosphere \citep{vernazza_1981}.  To heat the quiet corona takes about $2 \times 10^{21}$ W, and active areas of 30 times the quiet power per unit area \citep{klimchuk_2006}.  Conservatively assuming 10\% active area, the required power is $8 \times 10^{21}$ W.  These two points are used to calibrate the expansion into the heliosphere.

Using $s_{ex} = s_{en} + $ (size of region) and the size of the chromosphere as 2.5 Mm \citep{williams_2016} and assuming the voltage drop in the GMS is small, Eq. \ref{equation_6} gives the power deposited in the chromosphere in one hemisphere as:

\begin{equation}
1.325 \times 10^{22}\ \mathrm{W} = \frac{\omega \psi^2}{2f \mu_0} \left( \frac{1}{s_{en}} - \frac{1}{s_{en} + 2.5\  \mathrm{ Mm}} \right)
\label{equation_9}
\end{equation}

From cosmic ray attenuation, $\psi$ equals $3 \times 10^{14}$ Wb \citep{jiang_2011}.  This yields $s_{en}=1.5$ Mm, as shown in Fig.~\ref{figure_6}a) and $f=3.4$.  Thus inside the Sun, the solar dynamo must be thin if it powers the chromosphere. Using Eq. \ref{equation_5} and using a dynamo power from one hemisphere $P_{dym}=6\times10^{22}$ W then $s_{\rm inj}=0.53$ Mm, which is a reasonable value.  The validation is from the agreement with torsional oscillations, with the fast solar wind and with the current sheet discussed below.

The power to sustain half the GMS is about $13 \times 10^{20}$ W (assuming $B_o$ is 0.57 T) and lowers $P_{dym}$ by an amount much less than the uncertainty in its value.

The currents parallel to the expanding polar flux of the northern and southern hemispheres, which drives the slow solar wind, returns at the equatorial plane as the well-known toroidally-symmetric current sheet, which carries the current in and out of the equator, contributing to the corona on the 11-year cycle. The drop in current ($=I s_{\rm inj}/s$) with distance from the GMS produces a current across the magnetic field shown as dotted green lines in figure \ref{figure_6}b).

This current in the heliosphere produces a toroidal magnetic field between the GMS and the return current path. In the polar-flux-free region, this toroidal field pushes the plasma away from the sun. The geometry is that of an inverse-pinch \citep{anderson_1958}. The good curvature of the inverse pinch is stabilizing and it symmetrically expands out into space, which would suggest plasma from the photosphere is stably lifted off the Sun and accelerated by the inverse pinch effect, giving the slow solar wind near the equator. This inverse pinch effect may be the most effective at solar minima, also contributing to the maximum solar wind at these times, as observed. 

The current sheet reaches Earth with a predicted current of $2Is_{\rm inj}/s=10$ GA ($s$ is now the radius of Earth's orbit and 
$I=1.41 \times 10^{15}$ A 
is the current of one hemisphere).  At Earth, the current sheet is measured \citep{smith_2001} to be a 1D MECH state of $\pi$ radians with a lobe height of $10^7$ m and $B_o$ of 6 nT.  Thus, at Earth, this gives $\lambda_{HCS}=3.14 \times 10^{-7}$ m$^{-1}$.   From $\lambda=\mu_0j/B$ comes $j_{HCS}=1.5$ nAm$^{-2}$.  Taking the average radial component to be $0.7j_{HCS}$ yields a current of 10 GA.  This value agrees with that predicted, which is significant considering the current-sheet current at Earth is over five orders less than $I$.
This scaling is only valid in the slow solar wind that is a free expansion.

Torsional oscillations are speed-up and slow-down bands of the azimuthal flow that correlated with the solar cycle \citep{howe_2009}. The bands have a depth of $0.1R_o$ and solar activity is in the bands with positive acceleration. The oscillations are strongest near the surface and exist during solar minimum strongly confirming the model presented. Assuming that thermal convection distributes the torque from the dynamo down to the band depth, the flow speeds are consistent with the torque produced by the thin dynamo. When the current crosses the flux, a torque equal to $I \psi/4\pi$ is produced. The dynamo current can cross the flux many times with most of the net torque not escaping the Sun. The crossings with the dynamo drive are below the surface and take energy out of the rotation. Thus, the gray areas in Fig.~\ref{figure_5}a) have negative acceleration in Fig.~\ref{figure_5}b). Note the single-hemisphere dynamo of Fig.~\ref{figure_3}e) is confirmed by the gray area in Fig.~\ref{figure_5}a) and negative acceleration in Fig.~\ref{figure_5}b).

The crossing in the direction of the electric field are powering the heliosphere where field lines penetrate the surface, resulting in positive acceleration. Thus the blue and yellow areas of Fig.~\ref{figure_5}a) have positive acceleration in Fig.~\ref{figure_5}b). The expansion of the torque into the heliosphere is distributed in the same manner as is the power. 

To drive the 3 nHz amplitude of the torsional oscillations each torque must change the frequency by 6 nHz. A torque of $I \psi /4\pi$ will change the surface frequency of $\frac{3}{4}$ of the mass of a hemisphere between $0.9R_o$ and $1.0R_o$, at a radius of $0.9R_o$ by 6 nHz in 11 years. (The GMS covers about $\frac{3}{4}$ of the area.) These data strongly suggest the torsional oscillations are caused by the shallow, thin dynamo. 

The heliosphere is much lower density then under the solar surface. The current paths driving the fast solar wind are shown in Fig.~\ref{figure_6}. Where polar flux exists, the powerful dynamo torque quickly accelerates the heliosphere until the back EMF equals the highest dynamo driven EMF, giving the same rotation rate everywhere in the heliosphere as the fastest solar dynamo surface. The applied accelerating-torque, forces a nearly rigid rotation of the heliosphere (measured out to 15~$R_o$) \citep{lewis_1999} at near the frequency defined by the solar surface at the equator. This causes the fast solar wind. The dynamo torque rotates the heliosphere as a rigid body, even beyond the earth. A simple calculation shows the centrifugal force adds a radial velocity about equal to the tangential velocity. This equality of tangent and radial velocity in the solar wind is observed \citep{podesta_2008}. The sidereal solar wind speed is measured at $R_{sw}$ = 1 au \citep{podesta_2008} and at $R_{sw}$ = 1.4 au \citep{phillips_1995} to be approximately $\sqrt{1/2} \omega_{\rm solequ} R_{sw}$, where $\omega_{\rm solequ}$ is the angular frequency of the solar surface at the equator and $R_{sw}$ is the distance from the Sun. In the simplest terms, the solar dynamo forces the heliosphere to rotate with the Sun, giving the fast solar wind. If this equation where true at $R_{sw}$ = 54 au the fast solar-wind protons would be 2.4~MeV and the dynamo would be at full power, which is possible. 

Studying soft-x-rays from the corona \citep{chandra_2010} confirm details of the shallow dynamo. For example, in 1999 (Fig.~\ref{figure_2}b)) the equator driven dynamo is limited in latitude and does not dominate the entire Sun as it does in 1996--7.  Thus in 1999 the upper-latitude rotation of the corona is set more by the upper-latitude surface rotation and is slower. See Figure 2 of \citet{chandra_2010}.





Assuming that the polar flux is no thicker than $s_{\rm inj}/4$ then $B_o $is at least 0.57 T. This yields a 9-day toroidal Alfven time at the density of the Sun 1.8~Mm below the surface. The pressure equals the magnetic pressure at about 0.7~Mm below the surface.

\section{Results: flares, supergranules, sunspots, CMEs and prominences}

Flares are likely generated by similar physics as are on display at the poles during minima. The discussion of current drive so far has been for the solar minimum time, as in Fig.~\ref{figure_2}a). When the Sun is active, as in Fig.~\ref{figure_2}b), every field line that enters the Sun at a different latitude than it exits the Sun, will have current drive and power by the same mechanism. Figure~\ref{figure_2}b) shows the active Sun with current penetration over the whole surface, while the quiet Sun has only penetrations more-or-less confined to the equatorial regions and the polar regions. Flares are seen in the corona and the observed \citep{platten_2014} distribution of flares over the solar cycle agrees with the flux penetration distribution of the model. Thus, where flux is penetrating the solar surface, the dynamo power is released to the solar atmosphere as a flare.

The growth and decay times ($\approx$ 1 day) of the supergranules are about the meridian Alfven time and they appear to be the signature of relaxation-instability. One argument in favor of the GMS causing supergranulations is their pair-correlated movement up to 600 Mm \citep{hirzberger_2008}. This correlation on the solar radius scale is almost certainly due to magnetic effects. During solar minimum the GMS is sustained by the dynamo-driven polar flux requiring an orthogonal perturbation. A spheromak \citep{jarboe_1994} in a tuna-can shaped boundary of equal radius and height has a $\lambda_{eq}$ = 5/h, where h is the height.  This is smaller than that of the GMS, which is 2$\pi$/h, where h is the height of the plasma volume. As the spheromak increases in radius, $\lambda_{eq}$  approaches $\pi$/h. (It is within 3\% at radius of 5h.) Similarly, a hole through the center does not change $\lambda_{eq}$  significantly and thermal convective flow can freely pass through these regions.  This would explain the circular shape and flow through the supergranules. Thus the supergranules are probably spheromaks that are the orthogonal perturbations needed for the stable sustainment of the GMS by the dynamo-driven polar flux. 

The spheromak is a lowest eigenstate for the plasma volume and, therefore, the unsustained plasma relaxes to this state. However, it is incompatible with the injector geometry for sustainment and equilibrium and decays as fast as it is formed. Evidently, the energy that would be required to bend the polar flux around the spheromak so that it could sustain the spheromak, prevents this from being the sustained MECH state. Therefore, the MECH state sustained by the dynamo-driven polar flux is the GMS described.  However, the supergranulation has the ideal geometry for the required orthogonal perturbation. (For resistive MHD the perturbations must have a component of $\delta {\bf{v}} \times \delta {\bf{B}}$ parallel to ${\bf{B}}$ to sustain a MECH state.) Alternatively, the spheromak has some magnetic fields that are directed oppositely to the polar flux and some that are directed oppositely to the GMS fields. When the oppositely-directed fields reconnect, the spheromak connects the polar flux to the GMS flux and helicity will flow from the polar flux to the GMS. This anti-current drive in the polar flux and current drive in the GMS sustains the GMS.

\begin{figure}
\begin{center}
\includegraphics[width=\linewidth]{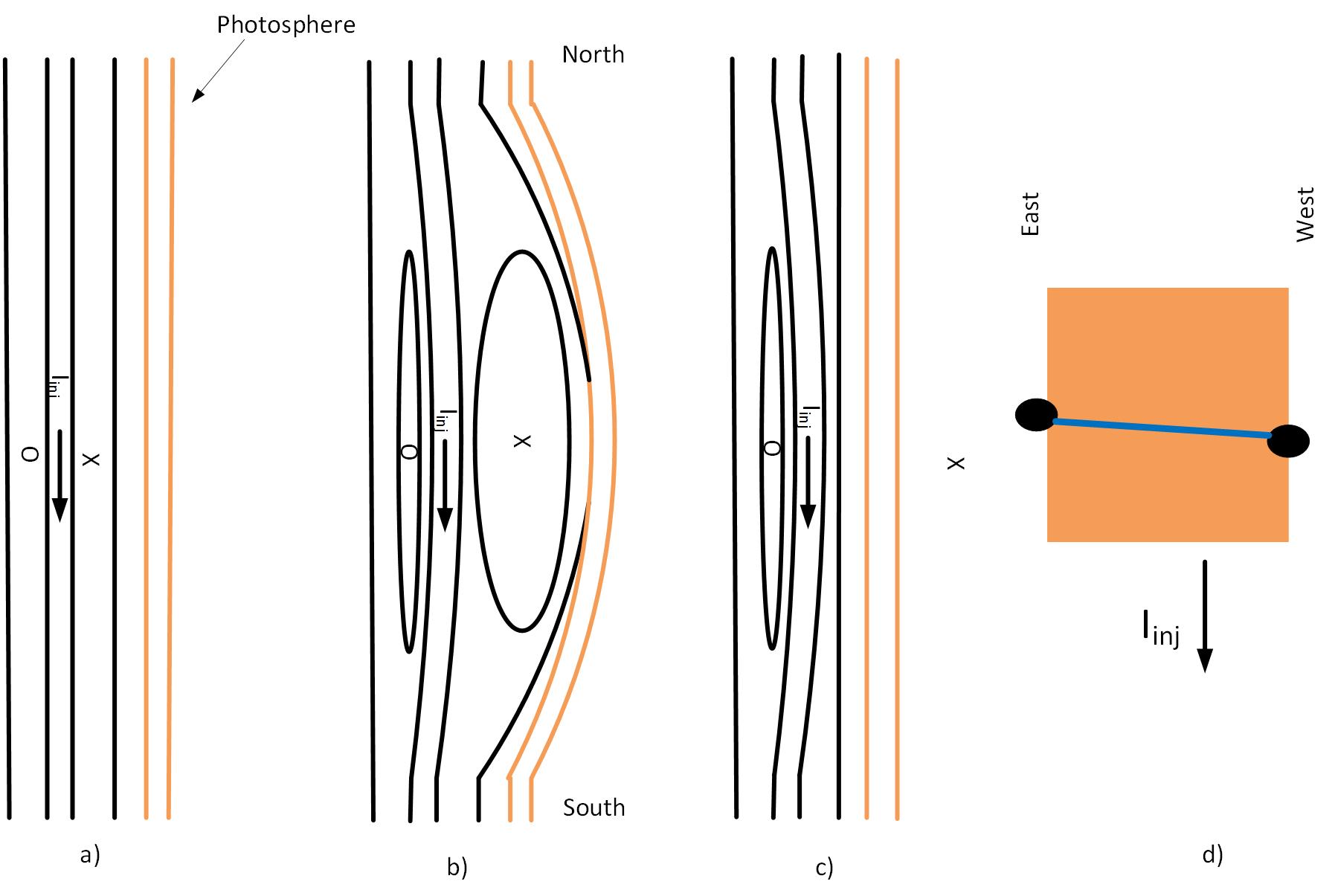}
\caption{Sketch of sunspot formation. This is a cross section of the meridional plane at the center of the 40 Mm square patch of the Sun being considered. $I_{\rm inj}$ is the dynamo driven current and on the current channel. The x and O represent flux going in and out of the page. a) The undisturbed GMS and photosphere. b) The locally thicker GMS with a lower lambda. c) After the sunspot pair is formed. This plane is between the sunspots. d) Top view of c), showing the sunspot-pair at the edges of the patch connected by the insulating break that allows the flux to escape. The toroidal flux of the sunspots is anchored by the polar flux that threaded the flat torus.}
\label{figure_7}
\end{center}
\end{figure}

Sunspots have too much flux to be formed from the local magnetic fields of the thin GMS.  They are more likely simply formed in place by helicity injection from the solar dynamo and the rest of the GMS. As the GMS moves to the surface, mass above the GMS becomes too small and is pushed away, and the GMS becomes locally thicker, as in Fig.~\ref{figure_7}b). In this patch of the GMS the polar-flux channel expands at constant $\lambda_{\rm inj}$ by the inflow of flux and current that link the polar flux channel, causing the flux to reverse and producing a flat torus that is threaded by the polar flux under the patch. Since the energetics are favorable, the flux linking the injector flux condenses to this volume. Thus an asymmetric flat torus, threaded by the polar flux of the patch, is formed. 

The gold lines in Fig.~\ref{figure_7} represent the very resistive part of the photosphere. When this region bounds the flat torus, it no longer confines the magnetic field of the upper thick area of the flat torus. Faculae could be arcing where the flux conserver is failing. The poloidal fields completely dissipate, and the toroidal field escapes but still enters and leaves the solar surface at the location of the resistive gap on the east and west edges of the 40 Mm patch. If the dynamo current under the patch is disrupted all of the flux linking this current becomes available to the sunspot. The toroidal flux ($\approx (\text{the poloidal flux)}/2\pi$) is $\approx 3 \times 10^{13}$ Wb in the flat torus, then a sunspot with a 0.2 T magnetic field and a diameter of 13 Mm is possible.

Thus the sunspot is formed at the footprints of the escaped toroidal flux. The angle of sunspots can vary because the angle of the resistive gap can vary. The toroidal field is not visible because it was stripped of its plasma as it passed through the insulating photosphere. The plasma expansion from venting may be observed as the Evershed clouds \citep{solana_2006}. They have a speed like the sound speed of the surface plasma; they vent at the edge of the flux and in the direction of the flux there (horizontal) as might be expected. This venting would cause expansion cooling of the region viewed through the sunspot. The sunspot lasts until the old polar flux that is trapping it leaves the Sun, giving a large variation in the life time.

(Thus, the sunspot magnetic field tends to be perpendicular to the internal dynamo current. Joy’s law \citep{hale_1919} for sunspots may occur because at first, near 30$\degree$ latitude, the actual internal polar flux is at an angle to the meridian direction.) 

If the flat torus also links some of the lower lobe’s flux and is strong enough to pull this lobe flux and the polar flux out of the surface as perhaps a CME then something like the transition from Fig.~\ref{figure_3}b) to Fig.~\ref{figure_3}c) would occur, producing a flare. The magnetic fields in CMEs have this structure \citep{rakowski_2011} and CME and flares tend to appear from the same event. If the flat torus is formed well below the surface it should tilt 90$\degree$ and form a short live spheromak.

\section{Discussion}

The values of $s_{en}$, $f$, $I$, and $s_{\rm inj}$ found in calibrating the expansion scaling using the power to the Chromosphere, power to the corona, $\psi$, and the total power are reasonable.
The expansion scaling accurately predicts the current sheet current at the earth and the velocity contour shape for torsional oscillations within the accuracy they are known.
The toroidal Alfven time leads to a reasonable estimate of the depth of the GMS.
The model also agrees with the detailed magnetic activity pattern of the solar cycle.

The helicity of two linked flux tubes is twice the product of the fluxes.
From Faraday’s law, the rate of change of flux linking a flux tube is the voltage along the flux tube.
With zero voltage defined as the equator and the solar-surface side of the polar flux and with the flux $\psi$ on the solar-surface side and zero on the other side of the polar flux the helicity injection rate is $2 \int \omega \left( \psi - \psi^\prime \right) d \psi^\prime / 2 \pi = \psi^2 \omega / 2\pi = V \psi$.
The voltage at the exit of the GMS is only slightly smaller than this because of the small power required by the GMS.
The expansion out of and away from the Sun is so fast compared to the magnetic diffusion time that the expanding toroidal flux links the same polar flux during the expansion.
Thus, most of the helicity created by the Sun flows away from the Sun with a small amount dissipated in the GMS and $V$ is approximately a constant of the expansion.
For the thin shallow GMS, the GMS magnetic fields of each hemisphere do not mix and the MECH state as discussed is expected.
The hemispheres have the opposite sign of helicity with zero total helicity.
Thus for a larger deeper dynamo the magnetic fields of the two hemispheres could, conceivably, mix and the zero net helicity would not prevent the loss of the magnetic fields much more rapidly than diffusion.

The polar flux flipping occurs by release of the old poloidal flux from the Sun and the emergence of the new anchored flipped flux from the solar surface, releasing magnetic energy, as discussed above.
The hot coronal plasma tied to the Sun by the old flux is then released to expand away from the Sun, also as a coronal mass ejection.
A new prediction of the proposed model is the release of anchoring plasma as part of the magnetic phenomena.

Finally, the thin sheets of magnetic plasma that form solar prominences are probably from the escaping upper lobe which is such a sheet.
The thickness of the GMS compared to its other dimensions is about like a sheet of paper.

\section{Conclusion}

The solar dynamo and the solar Global internal Magnetic Structure (GMS) appear to be a thin ($\sim 2$~Mm thick) structure near ($\sim 1$~Mm below) the solar surface.
Evidence for both properties is that the amplitude of the torsional oscillations are maximum at the surface within the resolution (1--2~Mm) of the measurement.
The magnitude of the current the dynamo drives parallel to the polar flux is proportional to the inverse thickness of the dynamo.
The evidence for the dynamo thickness is in agreement (to within the uncertainty of the measurement) with the measured power to the chromosphere; with measured power to the corona and the solar wind; with the current in the helio-current-sheet measured at the radius of the orbit of Earth;
with its forcing a rigid rotation, at the frequency of the solar surface at the equator, of the heliosphere out to at least 1.4 au, resulting in the fast solar wind;
and with the torque it supplies to drive the torsional oscillations.
Evidence for nearness to the surface is the size ($\sim 1$~Mm) of the expanding polar flux when it enters the photosphere, the observation that solar magnetic activity is generated near the surface and it is of the right order for a skin depth of the 22-year solar cycle.
That there is much evidence for, and little against, leaves very little doubt that the internal solar magnetic structure is a thin shallow GMS.

The thin stable magnetic equilibrium seems to be covering the solar surface just below the photosphere from 15 degrees, or less, to about 60 degrees latitude.
The magnetic field lines should be parallel to the solar surface and rotate with distance from the surface and to rotate $2 \pi$ radians in $\sim 2$~Mm.
This is expected for the sustained configuration that has the minimum magnetic energy for the helicity content, in accordance to the well-established minimum energy principle of plasma self-organization.
Resistive diffusion may help to push the magnetic fields to the surface and the GMS seems to lose $\pi$ radians every 11 years, causing the observed $180^{\circ}$ flipping of the solar magnetic fields including the polar flux.
This structure of the GMS is confirmed by the relationship of the velocity contours of the torsional oscillations with magnetograms.
Further evidence for this GMS and its loss is that solar prominences are made of thin sheets of magnetized plasma like the thin sheets lost by the GMS.
This loss is also consistent with the butterfly pattern of the sunspots and with the differences observed between solar maximum and solar minimum in the corona.
This evidence leaves little doubt the GMS is a sustained minimum energy state.

The roughly $\pi$ radians halves of the GMS are separated by the polar flux, which has a parallel current driven by the solar dynamo.
Perturbations cause the driven current to drive the adjacent current of the GMS, sustaining the GMS.
The magnetic perturbations may come from a transient state of lower energy than the GMS when it is not sustained but a higher energy if it were sustained, so it appears and decays, giving the necessary magnetic perturbations.
The transient state would be the spheromak.
The spheromak has the topology of, and is compatible with, the plasma flow pattern measured on the supergranules, which are measured to exist in the GMS region.
The life times of the supergranules agree with that expected for these transient states.
Thus, it is quite plausible that transient spheromaks supply the perturbation for cross-field current drive, or connect polar flux with the magnetic field of the GMS for the cross-field drive that sustains the GMS.
The supergranules are the signature of these spheromaks.

For completeness, it is likely that when the mass above the GMS becomes too small to keep the GMS submerged that a sausage like bubble grows a flat torus with a toroidal flux that links some polar flux.
If the polar flux is strong enough it traps the toroidal flux as the flat torus breaks the surface preventing the toroidal flux from leaving with the rest of the torus, causing sunspots where the toroidal flux leaves and enters the solar surface.
If the toroidal flux of the flat torus is strong enough it can pull the polar flux out of the surface as it leaves creating a CME and a flare.
All of these objects can dissipate quickly by leaving the solar surface.

\begin{acknowledgements}
The authors wish to thank many solar scientists for the highest quality data and Dr. Greg Kopp for help in estimating the dynamo power from the total solar irradiance data. This work supported by the U.S. Department of Energy Office of Science, Office of Fusion Energy Sciences under Award No. DE-FG02-96ER54361. 
\end{acknowledgements}

\bibliographystyle{aa} 
\bibliography{Solar_Paper_jarboe-AandA31} 

\begin{thebibliography}{66}
\expandafter\ifx\csname natexlab\endcsname\relax\def\natexlab#1{#1}\fi

\bibitem[{{Alfv{\'e}n}(1942)}]{alfven_1942}
{Alfv{\'e}n}, H. 1942, \nat, 150, 405

\bibitem[{Amenomori {et~al.}(2013)Amenomori, Bi, Chen, Chen, Chen, Cui,
  Danzengluobu, Ding, Feng, Feng, Feng, Gou, Guo, Hakamada, He, He, Hibino,
  Hotta, Hu, Hu, Huang, Jia, Jiang, Kajino, Kasahara, Katayose, Kato, Kawata,
  Labaciren, Le, Li, Li, Li, Liu, Liu, Liu, Lu, Meng, Mizutani, Munakata,
  Nanjo, Nishizawa, Ohnishi, Ohta, Onuma, Ozawa, Qian, Qu, Saito, Saito,
  Sakata, Sako, Shao, Shibata, Shiomi, Shirai, Sugimoto, Takita, Tan, Tateyama,
  Torii, Tsuchiya, Udo, Wang, Wu, Xue, Yamamoto, Yang, Yasue, Yuan, Yuda, Zhai,
  Zhang, Zhang, Zhang, Zhang, Zhang, Zhang, Zhaxisangzhu, \&
  Zhou}]{amenomori_2013}
Amenomori, M., Bi, X.~J., Chen, D., {et~al.} 2013, Phys. Rev. Lett., 111,
  011101

\bibitem[{Anderson {et~al.}(1958)Anderson, Furth, Stone, \&
  Wright}]{anderson_1958}
Anderson, O.~A., Furth, H.~P., Stone, J.~M., \& Wright, R.~E. 1958, The Physics
  of Fluids, 1, 489

\bibitem[{{Babcock}(1953)}]{babcock_1953}
{Babcock}, H.~W. 1953, \apj, 118, 387

\bibitem[{{Babcock}(1961)}]{babcock_1961}
{Babcock}, H.~W. 1961, \apj, 133, 572

\bibitem[{Birch {et~al.}(2013)Birch, Braun, Leka, Barnes, \&
  Javornik}]{birch_2013}
Birch, A.~C., Braun, D.~C., Leka, K.~D., Barnes, G., \& Javornik, B. 2013, The
  Astrophysical Journal, 762, 131

\bibitem[{Brandenburg(2005)}]{brandenburg_2005}
Brandenburg, A. 2005, The Astrophysical Journal, 625, 539

\bibitem[{Chandra {et~al.}(2010)Chandra, Vats, \& Iyer}]{chandra_2010}
Chandra, S., Vats, H.~O., \& Iyer, K.~N. 2010, Monthly Notices of the Royal
  Astronomical Society, 407, 1108

\bibitem[{Charbonneau(2010)}]{charbonneau_2010}
Charbonneau, P. 2010, Living Reviews in Solar Physics, 7, 3

\bibitem[{Charbonneau(2014)}]{charbonneau_2014}
Charbonneau, P. 2014, Annual Review of Astronomy and Astrophysics, 52, 251

\bibitem[{Choudhuri {et~al.}(2007)Choudhuri, Chatterjee, \&
  Jiang}]{choudhuri_2007}
Choudhuri, A.~R., Chatterjee, P., \& Jiang, J. 2007, Phys. Rev. Lett., 98,
  131103

\bibitem[{{Choudhuri} {et~al.}(1995){Choudhuri}, {Schussler}, \&
  {Dikpati}}]{choudhuri_1995}
{Choudhuri}, A.~R., {Schussler}, M., \& {Dikpati}, M. 1995, \aap, 303, L29

\bibitem[{Christensen-Dalsgaard(2002)}]{christensen-dalsgaard_2002}
Christensen-Dalsgaard, J. 2002, Rev. Mod. Phys., 74, 1073

\bibitem[{Christensen-Dalsgaard {et~al.}(1996)Christensen-Dalsgaard, Dappen,
  Ajukov, Anderson, {et~al.}}]{christensen-dalsgaard_1996}
Christensen-Dalsgaard, J., Dappen, W., Ajukov, S.~V., Anderson, E.~R., {et~al.}
  1996, Science, 272, 1286

\bibitem[{Christensen-Dalsgaard \& Thompson(2007)}]{christensen-dalsgaard_2007}
Christensen-Dalsgaard, J. \& Thompson, M. 2007, Observational results and
  issues concerning the tachocline, ed. D.~Hughes, R.~Rosner, \& N.~Weiss
  (Cambridge University Press), 53--85

\bibitem[{Cowling(1933)}]{cowling_1933}
Cowling, T.~G. 1933, Month. Not. Roy. Astron. Soc., 94, 39

\bibitem[{Dikpati \& Charbonneau(1999)}]{dikpati_1999}
Dikpati, M. \& Charbonneau, P. 1999, Astrophys. J., 518, 508

\bibitem[{Edenstrasser \& Kassab(1995)}]{edenstrasser_1995}
Edenstrasser, J.~W. \& Kassab, M. M.~M. 1995, Phys. Plasmas, 2, 1206

\bibitem[{Elsasser(1946)}]{elsasser_1946}
Elsasser, W.~M. 1946, Phys. Rev., 69, 106

\bibitem[{Elsasser(1955)}]{elsasser_1955}
Elsasser, W.~M. 1955, Am. J. Phys., 23, 590

\bibitem[{Haber {et~al.}(2002)Haber, Hindman, Toomre, Bogart, Larsen, \&
  Hill}]{haber_2002}
Haber, D.~A., Hindman, B.~W., Toomre, J., {et~al.} 2002, The Astrophysical
  Journal, 570, 855

\bibitem[{Hale(1908)}]{hale_1908}
Hale, G.~E. 1908, Terrestrial Magnetism and Atmospheric Electricity, 13, 159

\bibitem[{{Hale} {et~al.}(1919){Hale}, {Ellerman}, {Nicholson}, \&
  {Joy}}]{hale_1919}
{Hale}, G.~E., {Ellerman}, F., {Nicholson}, S.~B., \& {Joy}, A.~H. 1919, \apj,
  49, 153

\bibitem[{Hathaway(2010)}]{hathaway_2010}
Hathaway, D.~H. 2010, Liv. Rev. Solar Phys., 7, 1

\bibitem[{Hindman {et~al.}(2004)Hindman, Gizon, Jr., Haber, \&
  Toomre}]{hindman_2004}
Hindman, B.~W., Gizon, L., Jr., T. L.~D., Haber, D.~A., \& Toomre, J. 2004,
  Astrophys. J., 613, 1253

\bibitem[{Hirzberger {et~al.}(2008)Hirzberger, Gizon, Solanki, \&
  Jr.}]{hirzberger_2008}
Hirzberger, J., Gizon, L., Solanki, S.~K., \& Jr., T. L.~D. 2008, Solar
  Physics, 251, 417

\bibitem[{Hossack {et~al.}(2017)Hossack, Sutherland, \& Jarboe}]{hossack_2017}
Hossack, A.~C., Sutherland, D.~A., \& Jarboe, T.~R. 2017, Physics of Plasmas,
  24, 020702

\bibitem[{Howe(2009)}]{howe_2009}
Howe, R. 2009, Living Reviews in Solar Physics, 6, 1

\bibitem[{Jarboe {et~al.}(2012)Jarboe, Victor, Nelson, Hansen, Akcay, Ennis,
  Hicks, Hossack, Marklin, \& Smith}]{jarboe_2012}
Jarboe, T., Victor, B., Nelson, B., {et~al.} 2012, Nuclear Fusion, 52, 083017

\bibitem[{Jarboe(1994)}]{jarboe_1994}
Jarboe, T.~R. 1994, Plasma Physics and Controlled Fusion, 36, 945

\bibitem[{Jarboe {et~al.}(2015)Jarboe, Nelson, \& Sutherland}]{jarboe_2015}
Jarboe, T.~R., Nelson, B.~A., \& Sutherland, D.~A. 2015, Physics of Plasmas,
  22, 072503

\bibitem[{Jiang {et~al.}(2011)Jiang, Cameron, Schmitt, \&
  Sch\"{u}ssler}]{jiang_2011}
Jiang, J., Cameron, R.~H., Schmitt, D., \& Sch\"{u}ssler, M. 2011, Astron.
  Astrophys., 528, A83

\bibitem[{Klimchuk(2006)}]{klimchuk_2006}
Klimchuk, J.~A. 2006, Solar Physics, 234, 41

\bibitem[{Kolesnikov \& Obukhov-Denisov(1962)}]{kolesnikov_1962}
Kolesnikov, V.~N. \& Obukhov-Denisov, V.~V. 1962, Soviet Physics JETP, 15, 1001

\bibitem[{{Kopp, Greg}(2016)}]{kopp_2016}
{Kopp, Greg}. 2016, J. Space Weather Space Clim., 6, A30

\bibitem[{Larmor(1919)}]{larmor_1919}
Larmor, J. 1919, Rep. Br. Assoc. Adv. Sci., 87, 159

\bibitem[{{Leighton}(1969)}]{leighton_1969}
{Leighton}, R.~B. 1969, \apj, 156, 1

\bibitem[{Lewis {et~al.}(1999)Lewis, Simnett, Brueckner, Howard, Lamy, \&
  Schwenn}]{lewis_1999}
Lewis, D., Simnett, G., Brueckner, G., {et~al.} 1999, Solar Physics, 184, 297

\bibitem[{Maunder(1904)}]{maunder_1904}
Maunder, E.~W. 1904, Monthly Notices of the Royal Astronomical Society, 64, 747

\bibitem[{Moffatt(1978)}]{moffatt_1978}
Moffatt, H.~K. 1978, Magnetic Field Generation in Electrically Conducting
  Fluids (Cambridge University Press)

\bibitem[{{Parker}(1955)}]{parker_1955}
{Parker}, E.~N. 1955, \apj, 122, 293

\bibitem[{{Parker}(1963)}]{parker_1963}
{Parker}, E.~N. 1963, \apjs, 8, 177

\bibitem[{Parker(2009)}]{parker_2009}
Parker, E.~N. 2009, Solar Magnetism: The State of Our Knowledge and Ignorance,
  ed. M.~J. Thompson, A.~Balogh, {et~al.} (New York, NY: Springer New York),
  15--24

\bibitem[{Phillips {et~al.}(1995)Phillips, Bame, Barnes, Barraclough, Feldman,
  Goldstein, Gosling, Hoogeveen, McComas, Neugebauer, \& Suess}]{phillips_1995}
Phillips, J.~L., Bame, S.~J., Barnes, A., {et~al.} 1995, Geophysical Research
  Letters, 22, 3301

\bibitem[{{Platten, S. J.} {et~al.}(2014){Platten, S. J.}, {Parnell, C. E.},
  {Haynes, A. L.}, {Priest, E. R.}, \& {Mackay, D. H.}}]{platten_2014}
{Platten, S. J.}, {Parnell, C. E.}, {Haynes, A. L.}, {Priest, E. R.}, \&
  {Mackay, D. H.} 2014, Astron. Astrophys., 565, A44

\bibitem[{Podesta {et~al.}(2008)Podesta, Galvin, \& Farrugia}]{podesta_2008}
Podesta, J.~J., Galvin, A.~B., \& Farrugia, C.~J. 2008, Journal of Geophysical
  Research: Space Physics, 113, n/a, a09104

\bibitem[{Priest {et~al.}(2002)Priest, Heyvaerts, \& Title}]{priest_2002}
Priest, E.~R., Heyvaerts, J.~F., \& Title, A.~M. 2002, The Astrophysical
  Journal, 576, 533

\bibitem[{{R{\"a}dler}(2014)}]{radler_2014}
{R{\"a}dler}, K.-H. 2014, Astron. Nachr., 335, 459

\bibitem[{Rakowski {et~al.}(2011)Rakowski, Laming, \& Lyutikov}]{rakowski_2011}
Rakowski, C.~E., Laming, J.~M., \& Lyutikov, M. 2011, The Astrophysical
  Journal, 730, 30

\bibitem[{Savage {et~al.}(1997)Savage, Steigerwald, \& Title}]{savage_1997}
Savage, D., Steigerwald, B., \& Title, A. 1997, SOLAR MYSTERY NEARS SOLUTION
  WITH DATA FROM SOHO SPACECRAFT, {NASA} Press Release 97-256

\bibitem[{Smith(2001)}]{smith_2001}
Smith, E.~J. 2001, Journal of Geophysical Research: Space Physics, 106, 15819

\bibitem[{Solana {et~al.}(2006)Solana, Rubio, Beck, \& del
  Toro~Iniesta}]{solana_2006}
Solana, D.~C., Rubio, L. R.~B., Beck, C., \& del Toro~Iniesta, J.~C. 2006, The
  Astrophysical Journal Letters, 649, L41

\bibitem[{Spitzer(1962)}]{spitzer_1962}
Spitzer, L. 1962, Physics of Fully Ionized Gases, 2$^{\rm nd}$ Edition (New
  York, NY: John Wiley \& Sons)

\bibitem[{Spruit(2003)}]{spruit_2003}
Spruit, H. 2003, Solar Physics, 213, 1

\bibitem[{Steenbeck {et~al.}(1966)Steenbeck, Krause, \&
  R\"{a}dler}]{steenbeck_1966}
Steenbeck, M., Krause, F., \& R\"{a}dler, K.-H. 1966, Z. Naturforsch, 21a, 369

\bibitem[{Stenflo(2015)}]{stenflo_2015}
Stenflo, J.~O. 2015, Space Science Reviews, 1

\bibitem[{Taylor(1986)}]{taylor_1986}
Taylor, J.~B. 1986, Rev. Mod. Phys., 58, 741

\bibitem[{Temmer {et~al.}(2007)Temmer, Vr{\v{s}}nak, \& Veronig}]{temmer_2007}
Temmer, M., Vr{\v{s}}nak, B., \& Veronig, A.~M. 2007, Solar Physics, 241, 371

\bibitem[{Thomson(1903)}]{thomson_1903}
Thomson, J.~J. 1903, Conduction of Electricity Through Gases (University Press)

\bibitem[{Tobias(2002)}]{tobias_2002}
Tobias, S.~M. 2002, Philosophical Transactions of the Royal Society of London
  A: Mathematical, Physical and Engineering Sciences, 360, 2741

\bibitem[{Verkhoglyadova {et~al.}(2011)Verkhoglyadova, Tsurutani, Mannucci,
  Mlynczak, Hunt, Komjathy, \& Runge}]{verkhoglyadova_2011}
Verkhoglyadova, O.~P., Tsurutani, B.~T., Mannucci, A.~J., {et~al.} 2011,
  Journal of Geophysical Research: Space Physics, 116, A09325

\bibitem[{{Vernazza} {et~al.}(1981){Vernazza}, {Avrett}, \&
  {Loeser}}]{vernazza_1981}
{Vernazza}, J.~E., {Avrett}, E.~H., \& {Loeser}, R. 1981, \apjs, 45, 635

\bibitem[{Victor {et~al.}(2014)Victor, Jarboe, Hansen, Akcay, Morgan, Hossack,
  \& Nelson}]{victor_2014}
Victor, B.~S., Jarboe, T.~R., Hansen, C.~J., {et~al.} 2014, Physics of Plasmas,
  21, 082504

\bibitem[{Williams(2016)}]{williams_2016}
Williams, D.~R. 2016, {NASA} Sun Fact Sheet,
  https://nssdc.gsfc.nasa.gov/planetary/factsheet/sunfact.html

\bibitem[{Woltjer(1958)}]{woltjer_1958}
Woltjer, L. 1958, Proceedings of the National Academy of Sciences of the United
  States of America, 44, 489

\bibitem[{Yamada {et~al.}(2010)Yamada, Kulsrud, \& Ji}]{yamada_2010}
Yamada, M., Kulsrud, R., \& Ji, H. 2010, Rev. Mod. Phys., 82, 603

\end{thebibliography}

\end{document}